\documentclass[preprint,aps]{revtex4}
\usepackage{graphicx}

\begin{document}

\title{Topological $p_{x}+ip_{y}$ Superfluid Phase
of a Dipolar Fermi Gas in a 2D Optical Lattice}

\author{Bo Liu, Lan Yin}
\email{yinlan@pku.edu.cn} \affiliation{School of Physics, Peking
University, Beijing 100871, China}
\date{\today}

\begin{abstract}
In a dipolar Fermi gas, the anisotropic interaction between electric
dipoles can be turned into an effectively attractive interaction in
the presence of a rotating electric field. We show that the
topological $p_{x}+ip_{y}$ superfluid phase can be realized in a
single-component dipolar Fermi gas trapped in a 2D square optical
lattice with this attractive interaction at low temperatures.  The
$p_{x}+ip_{y}$ superfluid state has potential applications for
topological quantum computing.  We obtain the phase diagram of this
system at zero temperature.  When the interaction is weak, the  p-wave
superfluid state is stable for most filling factors except near half
filling where phase separation takes place.  In the weak-coupling limit,
the phase-separation region vanishes.  When the interaction strength is
above a threshold, the system is phase separated for any filling factor
$0<n<1$.  The transition temperature of the $p_{x}+ip_{y}$ superfluid
state is estimated and the implication for experiments is discussed.
\end{abstract}
\maketitle

Unconventional quantum states of ultra-cold Fermi gases have
attracted a lot of attention in the recent theoretical and
experimental studies.  Among them,  one of the most desirable states
is the topological superfluid $p_{x}+ip_{y}$ phase in a
two-dimensional (2D) single-component fermi gas. The zero-energy
Majorana modes were predicted to appear inside vertex cores in this
p-wave state \cite{Stern}. Due to their insensitivity to local
perturbations, Majorana fermions are very appealing for applications
in topologically-protected quantum computation \cite{Nayak}.

The p-wave superfluid phase was predicted to appear near a p-wave
Feshback resonance in a Fermi gas \cite{Chin, Regal}. However the
lifetime of the Fermi gas is severely limited by three-body
collisions, and so far this superfluid regime has not been achieved
in experiments.  The successful creation of $^{40}$K$^{87}$Rb polar
molecules \cite{Ni} has provided another platform to study
unconventional quantum states of Fermi systems.  For dipoles aligned
parallel to the $z$-direction, a p-wave superfluid state with the dominant
$p_{z}$ symmetry was predicted in a three-dimensional dipolar fermi gas
\cite{You}. In a dipolar Fermi gas trapped in $x-y$ plane, a $p_{x}$
or $p_{y}$ superfluid state may appear if the angle between the
dipole direction and the $x-y$ plane is small enough
\cite{Bruun}.  The $p_{x}+ip_{y}$ superfluid phase was
predicted for a 2D dipolar Fermi gas with a circularly-polarized
microwave field near the resonant frequency \cite{Cooper}.  The
d-wave and extended s-wave superfluid phases were also predicted for
a similar setup in an optical lattice with a linearly-polarized
resonant microwave field \cite{Liu}.  In experiments, however, the
precise control of the microwave polarization has been difficult
\cite{Wang}.

Here we propose a different method to realize a stable
$p_{x}+ip_{y}$ topological superfluid phase in a dipolar Fermi gas.
For $^{40}$K$^{87}$Rb or other fermionic polar molecules trapped in
a 2D square optical lattice, the dipole direction can be fixed by
applying an electric field.  If the electric field makes a small angle
with the lattice plane and rotates fast around the axis normal to
plane, the time-averaged interaction between dipoles is
effectively attractive \cite{Giovanazzi}. The $p_{x}+ip_{y}$ topological
superfluid state can be created in this dipolar Fermi
gas at low temperatures due to the attractive interaction.  The long lifetime of
$^{40}$K$^{87}$Rb molecules has been achieved in the optical lattice \cite{Amodsen},
which can be very helpful to stabilize this p-wave superfluid state.

\begin{figure}[t]
\begin{center}
\includegraphics[width=8cm]{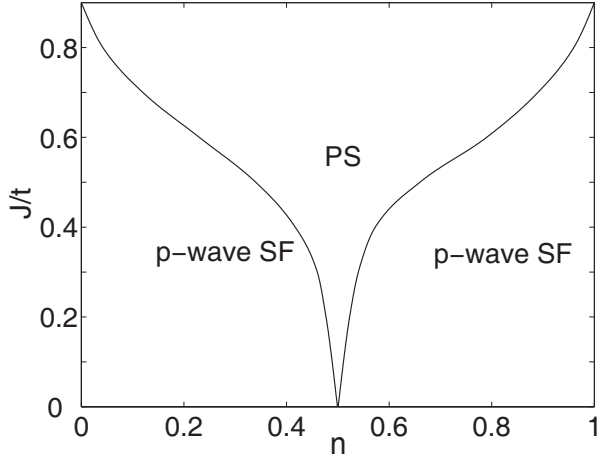}
\caption{Phase diagram at zero temperature.  The solid line is the
boundary between the $p_{x}+ip_{y}$ superfluid phase (p-wave SF) and
phase-separation region (PS).  The phase-separation region vanishes
in the weak coupling limit $J\rightarrow 0$.  When $J>0.89t$, the superfluid
phase vanishes.} \label{figure1}
\end{center}
\end{figure}

This work is organized as follows.  First we derive the effective
Hamiltonian for a single-component dipolar Fermi gas trapped in a
square lattice with a fast-rotating electric field.  Then we study
the $p_{x}+ip_{y}$ superfluid state in the self-consistent Hartree-Fock
approximation.  We obtain the phase diagram of this system at zero
temperature, as shown in Fig. \ref{figure1}.  When the interaction is weak,
the p-wave superfluid state is stable for most filling factors and
phase separation occurs close to half filling.  When the interaction
strength is above a critical value, the system is phase separated
for any $0<n<1$.  We also discuss the superfluid transition temperature
and its implication for experiments in the end.

{\it Effective model.} We consider a single-component dipolar Fermi
gas trapped in a square optical lattice under a rotating
electric field
\begin{equation}
{\bf E}(t)=E [\cos{\varphi} \hat{z}+\sin{\varphi}(\cos {\Omega}t
\hat{x}+\sin{\Omega}t \hat{y})],
\end{equation}
where $\Omega$ is the rotation frequency, $E$ is the magnitude of
the electric field, and $\varphi$ is the angle between the electric
field and $\hat{z}$ direction.  The optical lattice is in the $x-y$
plane and the rotation axis is along the $\hat{z}$ direction.  In
strong electric fields, dipoles are aligned parallel to ${\bf
E}(t)$.  For fast rotations, the effective interaction between
dipoles is the time-averaged interaction
\begin{eqnarray}
V_{dd}(r) &=&{d^2 (3\cos^{2}\varphi-1) \over 2 r^3},
\end{eqnarray}
where $d$ is the dipole moment, and $r$ is the distance between the
two dipoles. In the following we consider the case for
$\cos\varphi<\sqrt{1/3}$ where the effective interaction is
attractive, $V_{dd}(r)<0$.

The effective Hamiltonian of this dipolar Fermi gas is given by
\begin{equation}
H=-\sum_{<ij>}t(c_{i}^{\dagger}c_{j}+c_{j}^{\dagger}c_{i})+
\frac{1}{2}\sum_{i\neq j}V_{i-j}
c_{i}^{\dagger}c_{j}^{\dagger}c_{j}c_{i}, \label{hamiltonian 1}
\end{equation}
where $c_{i}$ is the fermion annihilation operator at lattice site
$i=(i_x, i_y)$, $V_{i-j}$ is the interaction between fermions at
site $i$ and $j$ given by $V_{i-j}=V_{dd}(|{\bf r}_i-{\bf r}_j|)$,
and ${\bf r}_i$ is the lattice vector. For simplicity we assume that
in the optical lattice the fermion can only hop between nearest
neighbors with hopping amplitude $t$. The interaction strength is
the largest for nearest neighbors, denoted by $J\equiv |V_{dd}(a)|=
d^2|3\cos^{2}\varphi-1|/(2a^3)$, where $a$ is the lattice constant.

The Hamiltonian in Eq. (\ref{hamiltonian 1}) has the important
particle-hole symmetry.  Under the transformation $c_{i}^{\dagger}
\rightarrow c_{i}$ and $c_{i} \rightarrow c_{i}^{\dagger}$, a state
with filling factor $n$ turns into another state with filling factor
$1-n$.  The chemical potential $\mu$ transforms as $\mu \rightarrow
-\mu+V(0)$ where $V(0)=\sum_{m \neq 0} V_{m}=-9.02J$.  Note that
the hopping amplitude $t$ becomes $-t$ under this transformation,
but in the square lattice this phase change can be eliminated by a
unitary transformation $c_{i} \rightarrow -c_{i}$ for only one
sublattice with the same odevity of $i_x+i_y$.

{\it Self-consistent Hartree-Fock approximation.} Due to the
attractive interaction, fermions tend to pair with each other at low
temperatures.  In the square lattice, since the interaction strength
is the largest between nearest neighbors, the pairing
probability is the largest for fermions from nearest neighbors.
Under the constraint of Fermi statistics, the dominant pairing
symmetry should be p-wave.  To study this superfluid pairing state,
we use the self-consistent Hartree-Fock approximation in which the
interaction produces the Hatree energy $E_h$, Fock energy $E_x$, and
the pairing energy $E_p$ given by
\begin{eqnarray}
E_h &=& \frac{1}{2}\sum_{i\neq j}V_{i-j}n^2, \nonumber \\
E_x &=& -\frac{1}{2}\sum_{i\neq j}V_{i-j}\langle c_{i}^{\dagger}c_{j}
\rangle \langle c_{j}^{\dagger}c_{i}\rangle, \nonumber \\
E_p &=& \frac{1}{2}\sum_{i\neq j}V_{i-j}\langle c_{i}^{\dagger}
c_{j}^{\dagger}\rangle \langle c_{j}c_{i}\rangle,
\end{eqnarray}
where $n=\langle c_{i}^{\dagger}c_{i}\rangle$ is the filling factor.
The Hamiltonian in Eq. (\ref{hamiltonian 1}) can be approximated as
\begin{widetext}
\begin{eqnarray}
H'&=&-\sum_{<ij>}t(c_{i}^{\dagger}c_{j}+h.c.)+\frac{1}{2}\sum_{i\neq j}
V_{i-j} [n ( c_{i}^{\dagger}c_{i}+c_{j}^{\dagger} c_{j}) \nonumber
-(\langle c_{i}^{\dagger}c_{j} \rangle c_{j}^{\dagger}c_{i}+h.c.)
+(\langle c_{i}^{\dagger}c_{j}^{\dagger} \rangle c_{j}c_{i}+h.c.)]-E_I,
\end{eqnarray}
\end{widetext}
where $E_I=E_h+E_x+E_p$.

In momentum space, we have
\begin{equation}
H'-\mu \hat{N}_f=\sum_{{\bf k}}[\xi_{\bf k}c_{{\bf
k}}^{\dagger}c_{{\bf k}}+{\Delta_{\bf k}^{*} \over 2} c_{-{\bf
k}}c_{{\bf k}}+{\Delta_{\bf k} \over 2}c_{{\bf k}}^{\dagger}c_{-{\bf
k}}^{\dagger}]-E_I, \label{MFSFHM}
\end{equation}
where $\xi_{\bf k}=\varepsilon_{\bf k}+\Sigma_{\bf k}-\mu$,
$\varepsilon_{\bf k}=-2t(\cos{k_xa}+\cos{k_ya})$ is the band energy,
$\mu$ is the chemical potential, $\Sigma_{\bf k}$ is the
Hartree-Fock self-energy given by $$\Sigma_{\bf k}=V(0)n-{1 \over N}
\sum_{{\bf k}^{\prime}}V({{\bf k}-{\bf k}^{\prime}})n_{{\bf
k}^{\prime}},$$ the pairing gap $\Delta_{\bf k}$ is given by
$$\Delta_{\bf k}={1 \over N}\sum_{{\bf k}^{\prime}}V({{\bf k}-{\bf
k}^{\prime}})g_{{\bf k}^{\prime}},$$ $N$ is the total number of
lattice sites and $\hat{N}_f$ is the fermion number operator,
$$V({\bf k})= \sum_{m\neq 0}V_{m}\exp(-i{\bf k}\cdot {\bf r}_m),$$
$g_{\bf k}=\langle c_{-{\bf k}}c_{{\bf k}}\rangle$, and $n_{\bf
k}=\langle c^\dagger_{\bf k}c_{\bf k}\rangle$.

The Hamiltonian in Eq. (\ref{MFSFHM}) can be diagonalized by
Bogoliubov transformation.  The Hartree-Fock self-energy and gap can
be determined self-consistently
\begin{eqnarray}
\Sigma_{\bf k} &=& V(0)n-{1 \over N}\sum_{{\bf k}^{\prime}}
V({{\bf k}-{\bf k}^{\prime}}){{1}\over{2}}[1-\frac{\xi_{{\bf k}^{\prime}}}
{E_{{\bf k}^{\prime}}} \tanh({\beta \over 2}E_{{\bf k}^{\prime}})], \nonumber \\
\label{SE 2}\\
\Delta_{\bf k} &=& -{1 \over N}\sum_{{\bf k}^{\prime}}V({{\bf k}-{\bf k}^{\prime}})
\frac{\Delta_{{\bf k}^{\prime}}}{2E_{{\bf k}^{\prime}}}\tanh({\beta \over 2}
E_{{\bf k}^{\prime}}), \label{GAP 2}
\end{eqnarray}
where the quasi-particle excitation energy $E_{\bf k}$ is given by
$E_{\bf k}=\sqrt{\xi_{\bf k}^2+|\Delta_{\bf k}|^2}$ and
$\beta=1/(k_BT)$.  The filling factor $n$ can be also determined
self-consistently,
\begin{equation}
n={{1}\over{2}}[1-{{1}\over{N}}\sum_{{\bf k}}
\frac{\xi_{\bf k}}{E_{\bf k}}\tanh({\beta \over 2}E_{{\bf k}})].
\label{number}
\end{equation}

\begin{figure}[t]
\begin{center}
\includegraphics[width=8cm]{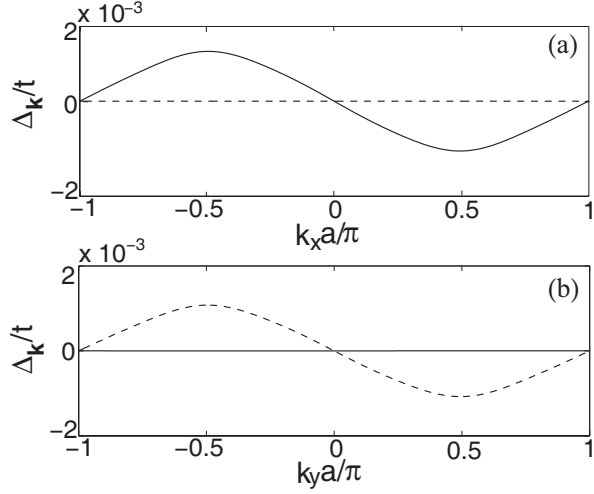}
\caption{The gap $\Delta_{\bf k}$ versus wavevector ${\bf k}$ at
$J/t=0.6$, $n=0.11$, and $T=0$.  The solid and dashed lines are the
real and imaginary parts of the gap. (a) At ${k_y}a/\pi=0$, the real
part of gap has the same sign structure as the function
$\sin({k_x}a)$, and the imaginary part vanishes.  (b) At
${k_x}a/\pi=0$, the imaginary part of gap has the same sign
structure as the function $\sin({k_y}a)$, and the real part
vanishes.}  \label{figure2}
\end{center}
\end{figure}

We numerically solve the Hartree-Fock self-energy equation (\ref{SE
2}), gap equation (\ref{GAP 2}), and number equation (\ref{number})
together, and find that the gap $\Delta_{\bf k}$ has finite
magnitude and displays a complex symmetry below a critical
temperature.  A global unitary transformation can always be applied
so that the real part of the order parameter $\Delta_{\bf k}$
remains the same under the transformation $k_y \rightarrow -k_y$ but
changes sign when $k_x \rightarrow -k_x$, while the imaginary part
of the gap transforms oppositely.  As shown in Fig. \ref{figure2}, we
find from numerical computation that the real part of the gap has
the same sign structure as the function $\sin({k_x}a)$ and the
imaginary part has the same sign structure as the function
$\sin({k_y}a)$.

\begin{figure}[t]
\begin{center}
\includegraphics[width=8cm]{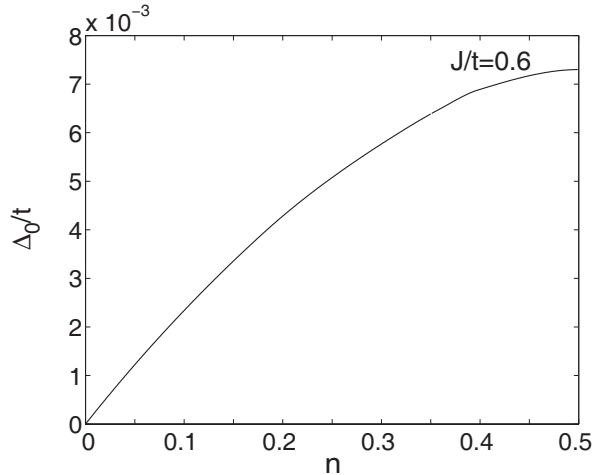}
\caption{The gap magnitude $\Delta_0$ versus filling factor $n$ for
$J/t=0.6$ and $T=0$, where $\Delta_0=|\Delta_{\bf k}|$ at ${\bf
k}=(\pi/2a, \pi/2a)$.  At half filling, $\Delta_0$ has the largest
value. The gaps for filling factors $0.5<n<1$ can be obtained from
particle-hole symmetry.}  \label{figure3}
\end{center}
\end{figure}

The magnitude of the gap is a monotonically increasing function of
the interaction strength $J$.  For the fixed interaction strength,
the magnitude of the gap also increases with the filling factor up
to half filling, as shown in Fig. \ref{figure3}.  Due to the
particle-hole symmetry, one state at filling $n=\delta$ is a dual to
another state at $n=1-\delta$, and they share the same order
parameter.  Thus the gap of the superfluid state above half filling
can always be obtained from the state below half filling.  At
half-filling, the magnitude of the gap is the largest.

\begin{figure}[t]
\begin{center}
\includegraphics[width=8cm]{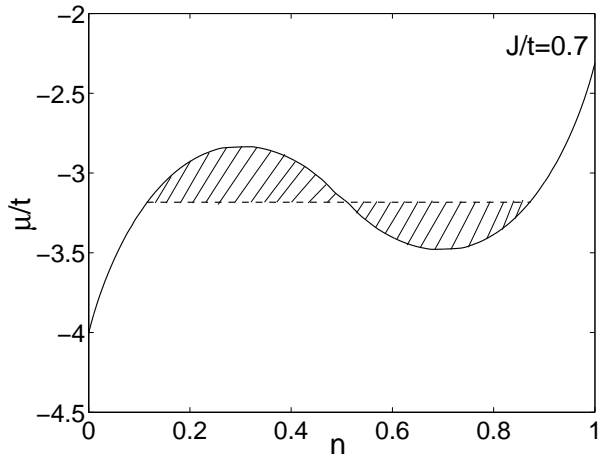}
\caption{Chemical potential $\mu$ versus the filling factor $n$ at
$J/t=0.7$ and $T=0$.  The dashed line between $n=0.12$ and $n=0.88$
shows the phase-separation region.  The shaded areas above and below
the dashed line are equal, as given by the phase-separation condition.}  \label{figure4}
\end{center}
\end{figure}

{\it Zero-Temperature phase diagram.} Due to the attractive nature
of the effective dipole-dipole interaction, the energy of this
dipolar Fermi gas is smaller than that of an ideal Fermi gas trapped
in the same optical lattice.  This energy reduction caused by the
interaction increases with the filling factor.  When the interaction
strength is large enough, the interaction effect is dominant and the
system can be unstable. As shown in Fig. \ref{figure4}, at $T=0$ and
$J/t=0.7$, for $0.3<n<0.7$, the chemical potential is a monotonically
decreasing function of the filling factor and the compressibility is
negative, indicating that the $p_{x}+ip_{y}$ superfluid state
is dynamically unstable.  The system is phase separated in the
unstable region and becomes a mixture of $n=0.12$ and $n=0.88$
states, shown by the dashed line in Fig. \ref{figure4}. Due to the
particle-hole symmetry, the phase-separation region is symmetric
with respect to the half-filling point.

We find that the phase-separation region increases with the
interaction strength.  In the weak-coupling limit, $J\rightarrow 0$,
the phase-separation region vanishes.  At a critical interaction strength,
phase separation occurs between $n=0$ and $n=1$, and
the chemical potential satisfies $\mu(0)=\mu(1)$, where $\mu(n)$ is
the chemical potential at filling factor $n$ and $\mu(0)=-4t$. Due
to the particle-hole symmetry, $\mu(0)+\mu(1)=V(0)$, this critical
interaction strength satisfies $V(0)=-8t$ and $J=0.89t$.  When the
interaction strength is above the critical value, $J>0.89t$, the
system becomes a mixture of $n=0$ and $n=1$ states which are insulating
states due to zero particle or hole densities.  Based on these
results, the zero-temperature phase diagram of this system is obtained
and shown in Fig. \ref{figure1}.

{\it Discussion and conclusion.} When the temperature increases, the
phase-separation region rapidly shrinks due to the increase in entropy.
For example, we find that at $k_BT=0.1t$ and $J=t$, the phase
separation nearly disappears and the superfluid phase exists even
close to half filling.  In the superfluid region, as the temperature
increases, this 2D dipolar Fermi gas eventually undergoes a
Berezinskii-Kosterlitz-Thouless (BKT) transition from the superfluid
state to the normal state.  At the critical point,
the vortex-antivortex pair disassociates
and it costs zero free energy to generate a single unbound vortex.

The transition temperature $T_{BKT}$ is given by
\cite{Berezinskii, Kosterlitz},
\begin{equation}
T_{BKT}={{\pi}\over{8 k_B}}\rho \label{BKT},
\end{equation}
where $\rho $ is the superfluid density.  In the $p_{x}+ip_{y}$
superfluid state, the superfluid density refers to the geometric
mean of eigenvalues of the superfluid density tensor which describes the
response of the system to phase twists in the superfluid order parameter.
At finite temperatures, the superfluid density is reduced due to
excitations of fermionic quasi-particles and bosonic (collective)
modes. Here, we assume that the bosonic excitations are negligible
and the superfluid density tensor is given by
\begin{equation}
\rho_{ij}={{1}\over{N a^2 }}\sum_{\bf {k}}(n_{\bf
{k}}\partial_i\partial_j\varepsilon_{\bf {k}}-Y_{\bf {k}}\partial_i\varepsilon_{\bf
{k}}\partial_j\varepsilon_{\bf {k}}) \label{SPD}
\end{equation}
where $\partial_i f \equiv \partial f/\partial k_i$ and $Y_{\bf
{k}}=\beta {\rm sech}^2(\beta E_{\bf k}/2)/4$ is the Yoshida
distribution.  Since the off-diagonal matrix elements of the
superfluid density tensor are small, the superfluid density is
approximately given by $\rho \approx (\rho_{xx}+\rho_{yy})/2$.

By solving Eq. (\ref{BKT}) and (\ref{SPD}) numerically, we obtain the BKT temperature
$k_BT_{BKT}=0.11 t$ for $J=t$ and $n=0.48$.  The lattice constant and potential strength of the
optical lattice can be tuned to increase the hopping
amplitude $t$ and interaction strength $J$, which can be helpful for
observing the $p_{x}+ip_{y}$ superfluid state in experiments.  In the current
experiment \cite{Amodsen} on KRb molecules, the lattice constant is given by $2a=1064$nm.
The hopping amplitude $t$ is about $0.2E_R$ for lattice
potential $V_L=10E_R$ \cite{Wall}, and the strength of the
effective dipole-dipole interaction can be tuned up to $0.2E_R$ for the
singlet rovibrational state of KRb molecules, where $E_R$ is the recoil energy.
Under these conditions $J=t=0.2E_R$ and $n=0.48$, the BKT temperature $T_{BKT}$ is approximately
2nK.  Similar estimate can be obtained for NaK
molecules which have a larger dipole moment of $2.7$D.  NaK molecules have no
reactive channels of decay and can be confined in a lattice with a smaller lattice
constant $2a\approx500$nm.  With a larger
recoil energy $E_R$, under the same condition $J=t=0.2E_R$ and $n=0.48$,
the BKT temperature $T_{BKT}$ of NaK molecules is
approximately $20$nK.

In summary, we propose that the topological $p_{x}+ip_{y}$
superfluid state can be created in a dipolar Fermi gas trapped in a
2D optical lattice in the presence of a rotating electric field, due
to the effectively attractive dipole-dipole interaction. The phase
diagram of this system is obtained at zero temperature.  For weak
interaction, the $p_{x}+ip_{y}$ superfluid state is stable for most
filling factors and phase separation occurs near half filling.  When
the interaction strength is above the critical value $J>0.89t$, the system
is phase separated between filling factors $n=0$ and $n=1$.  We estimate
the BKT transition temperature of the topological $p_{x}+ip_{y}$ superfluid
state which may be accessible in future experiments.  We would like to thank
T.-L. Ho for helpful discussions.  This work is supported by NSFC under Grant No
10974004.

\end{document}